\documentclass{icrc2009}

\usepackage{graphicx}   
\usepackage{caption}    
\usepackage[font=footnotesize]{subfig} 
\usepackage{fixltx2e}
\usepackage{url}

\newcommand{\shorttitle}[1]%
{\markboth{Proceedings of the 31\MakeLowercase{$^{st}$} ICRC, {\L}\'{o}d\'{z} 2009}{#1} }

\begin{document}
\title{Timing Calibration of the ANTARES Neutrino Telescope}

\author{\IEEEauthorblockN{Juan Pablo G\'omez-Gonz\'alez\IEEEauthorrefmark{1}\\
On behalf of the ANTARES Collaboration} \\
                            
\IEEEauthorblockA{\IEEEauthorrefmark{1} IFIC- Instituto de F\'isica Corpuscular, Edificios de Investigaci\'on de Paterna,\\ 
CSIC - Universitat de Val\`encia, Apdo. de Correos 22085, 46071 Valencia, Spain.}}

\shorttitle{Juan Pablo G\'omez-Gonz\'alez. Timing Calibration in ANTARES.}
\maketitle

\begin{abstract}
On May 2008 the ANTARES collaboration completed the installation of a neutrino telescope in the Mediterranean Sea. This detector consists of a tridimensional array of almost 900 photomultipliers (PMTs) distributed in 12 lines. These PMTs can collect the Cherenkov light emitted by the muons produced in the interaction of high energy cosmic neutrinos with the matter surrounding the detector. A good timing resolution is crucial in order to infer the neutrino track direction and to make astronomy. 
In this presentation I describe the time calibration systems of the ANTARES detector including some measurements 
(made both at the laboratory and in-situ) which validate the expected performance.  
\end{abstract}

 
\section{Introduction}
The ANTARES Collaboration has completed the construction of the largest underwater neutrino telescope in the Northern hemisphere [1].  The final detector, located 40 km off the Toulon coast (France) at 2475 m depth, consist of an array of 884 photomultipliers (PMT) distributed along 12 lines separated by $\sim$74 m. Each line has 25 floors (or storeys) holding triplets of 10$"$ PMTs housed in pressure resistant glass spheres called Optical Modules (OM). The clock signal, slow control commands, HV supply, and the readout, arrive at the OMs via the electronic boards housed in the Local Control Module (LCM) container made of titanium. The Bottom String Socket (BBS) anchors each line to the seabed while a buoy at the top gives them vertical support. The whole detector is operated from a control room (shore station) via the electro-optical cable, linked to the Junction Box (JB) which splits the connection to the BBS of the 12 lines (see Fig. 1).
 
The aim of the experiment is to detect high energy cosmic neutrinos and to identify the source of emission by reconstructing the muon tracks. Since the track reconstruction algorithms are based on the arrival time of the Cherenkov photons in the PMTs and the position of the PMTs, both  timing and positioning in-situ  calibration are needed. In ANTARES we have developed several timing  and positioning calibration systems to ensure the expected performance, i.e., an angular resolution  better than 0.3 deg at energies larger than 10 TeV. This paper describes the timing calibration system of  ANTARES while details on the positioning system can be found in [2].\\

 \begin{figure}[!t]
  \centering
  \includegraphics[width=3.2in]{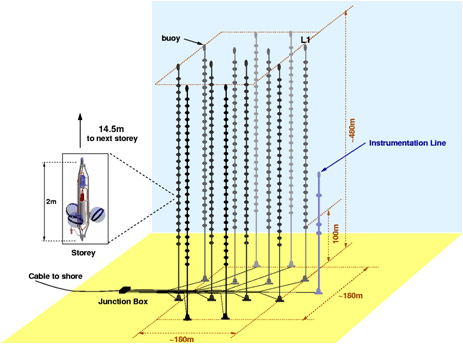}
  \caption{Schematic view of the ANTARES detector layout. The active part are the photomultiplier tubes grouped in triplets in each storey.}
  \label{simp_fig}
 \end{figure}
 
\section{Timing Calibration}
Concerning to the time calibration we can distinguish between absolute and relative time resolution. The requirement on absolute temporal resolution has the main purpose of correlate detected signals with astrophysical transient events (GRBs, AGN flares, SGR bursts).
In this case the precision needed to measure the time of each event with respect to the universal time (to establish correlations) is of the order of milliseconds. The absolute timestamping  is  made by interfacing the shore station master clock and a card receiving the GPS time ($\sim$100 ns accuracy with respect to UTC). The main uncertainty comes from the electronic path between the electro-optical cable that links the junction box and the shore station.

The relative time resolution refers to the individual time offsets in the photon detection due to differences on the transit times and the front-end electronics among PMTs. The main uncertainties come from the transit time spread (TTS) of the signal in the PMTs ($\sim$1.3 ns) and the optical properties of the sea water (light scattering and chromatic dispersion), which means $\sim$1.5 ns considering 40 m distance.  Therefore, all electronics and calibration system are required to contribute less than 0.5 ns to the relative time resolution in order to guarantee the expected angular resolution. 

\section{Calibration systems}
Several techniques have been developed in ANTARES in order to perform the time calibration [3] of the detector: (A)
The on-shore calibration is performed at laboratory to check the individual time offsets and calibrate the electronics before the deployment in the sea. (B) The clock system enables the measurement of the signal time delay from the clock board located on each OM to the shore station. (C) The internal LED monitors the PMT transit time. (D) The Optical Beacons system, which consist of LED and laser sources of pulsed light with a well know emission time, are used for in-situ timing calibration. (E) The K40 calibration allows (together with the Beacons) the determination of the time differences among PMTs in the same floor, which is used as a check of the time offsets previously determined.
The main features of these method are summarized in this section.

\subsection{On-shore calibration systems}
The on-shore calibration is necessary in order to check that all detector components work properly before the deployment in the sea site. Moreover, they allow us to obtain the first calibration parameters wich will be confirmed and sometimes corrected by the in-situ systems. To this end each integration site of ANTARES has a setup consisting in a laser ($\lambda$ = 532 nm) sending light through an optical fiber to the PMTs and to the OBs placed in a dedicated dark room. Knowing the difference between the time emission of the laser light with respect to the time when the signal is recorded by the PMT we can compute the individual PMT offsets after correcting by the time delay of the fiber light transport. Taking one PMT as reference the relative time offsets can be corrected for the whole detector.

\subsection{Clock system}
The clock system consists of a 20 MHz clock generator on shore synchronized with the GPS, plus a clock distribution system and a signal transceiver on each LCM. The aim is to provide a common signal to all the PMTs. It works essentially by sending optical signals  from shore to each LCM, where they are sent back as soon as they arrive. The corresponding round trip time is twice the time delay due to the cables length for each individual LCM. The uncertainties when computing the time difference between the anchor of a line and one particular floor are of the order of 15 ps, good enough for our purposes in relative time calibration. Considering the time difference between the shore station and the JB we found that the fluctuations in the whole trip delay are around 200 ps, which fulfills the requirements for the absolute time calibration.

\subsection{Internal LED}
An internal blue ($\lambda$ = 472 nm) LED is glued to the back of every PMT in order to illuminate the photocathode from inside. The aim of this device is to monitor the transit time (TT) of the PMT measuring the difference between the time of the LED flash and the time when the flash light is recorded by the PMT. The results obtained with this system indicate that the average TT of the PMTs is stable within 0.5 ns.\\

	\begin{figure}[!t]
   \centering
   \includegraphics[width=3.0in]{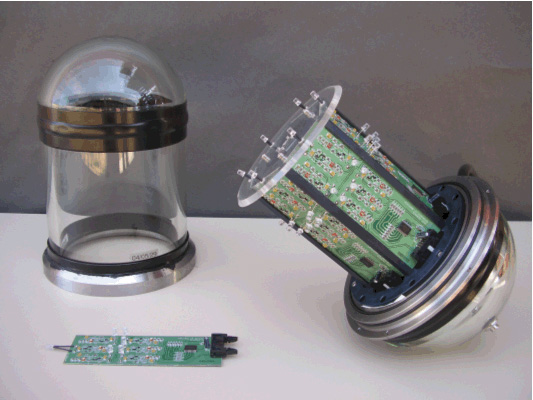}
   \caption{The LED Optical Beacon without the upper cap of it borosilicate container. The light emission time is measured with an  internal PMT. One of the board circuits is also shown separately. }
   \label{simp_fig}
  \end{figure}

\subsection{Optical Beacons}
The in-situ calibration of the time offsets is performed with a system of external light emitters called Optical Beacons (OBs) [4]. This system comprises two kinds of devices: (1) LED Optical Beacons, and (2) Laser Beacon.  LED Beacons are conceived to calibrate the relative time offsets among OMs within each line. Moreover they can also be used with other purposes such as measuring and monitoring optical water properties, or studying possible PMT efficiency loss. Laser Beacon is being used for interline calibration and cross-checking of the time offsets of the lower floors.\\

\begin{figure}[!t]
  \centering
  \includegraphics[width=2.5in]{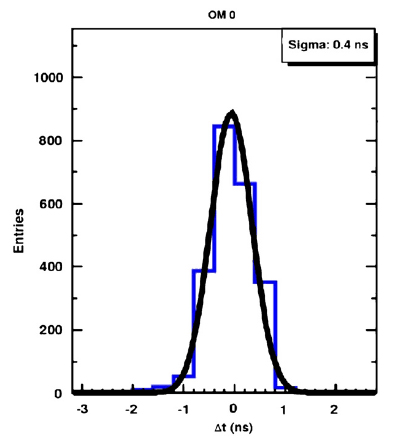}
  \caption{Time difference distribution between a PMT and a LED OB placed 14.5 m away. The standard deviation of 0.4 ns can be understood as an estimation of the ANTARES time resolution from the electronics. }
  \label{simp_fig}
 \end{figure} 

\begin{enumerate}
	\item The LED Optical Beacon (LOB) is a device made up by 36 individual blue LEDs ($\lambda$ = 472 nm) arranged in groups of six on six vertical boards which are placed side by side forming an hexagonal cylinder (see Fig. 2).  The 36 LEDs in the Beacon can be flashed independently or in combination, and at different intensities (maximum intensity produces 160 pJ per pulse). In ANTARES each standard line contains four LOBs placed every 2, 9, 15 and 21 floors in order to illuminate the OMs located above in the same line.
The time offset computation is based on the time residuals defined as the difference between the time emission of the LED light with respect to the time when the flash is recorded by the OM. In order to know the time emission of the LED light, a small photocathode is placed inside the frame of the LOB. To check the validity of a set of time offset we have calculated the time difference by pairs of OMs in the same storey. The results obtained from the in-situ computation have shown that the changes of these offsets with respect to the measured values in the integration sites are not very large in general, and that only 15\% of the PMTs need a correction greater than 1 ns. The LOBs have also been used to check the ANTARES time resolution. One way to do that is by flashing nearby PMTs. In this case, the contribution of the TTS, which is inversely proportional to the square root of the number photo-electrons, is negligible due to the great amount of light collected by the PMT. The contribution of the photon dispersion is small, since the arrival time distribution is dominated by the first photons.  Finally, the contribution from the small internal OB PMT is also negligible. Therefore, one can safely say that the only important contribution comes from the electronics. We have found that this term takes values of around 0.4 ns (see Fig. 3), which is lower than the 0.5 ns required for the relative time calibration in order to reach the expected angular resolution.\\

 \begin{figure}[!t]
   \centering
   \includegraphics[width=3.0in]{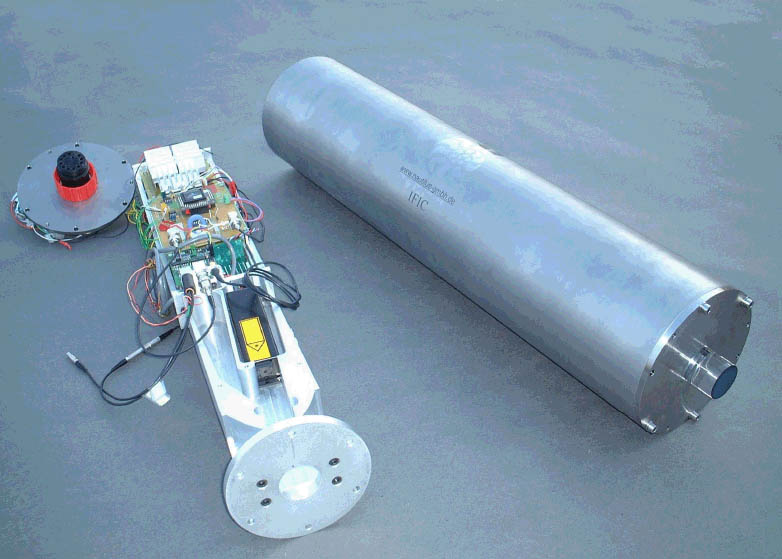}
   \caption{The LED Optical Beacon with it titanium container. The inner mechanics holding the laser head and its associated electronics are visible.}
   \label{simp_fig}
 \end{figure}

	\begin{figure}[!t]
   \centering
   \includegraphics[width=2.5in]{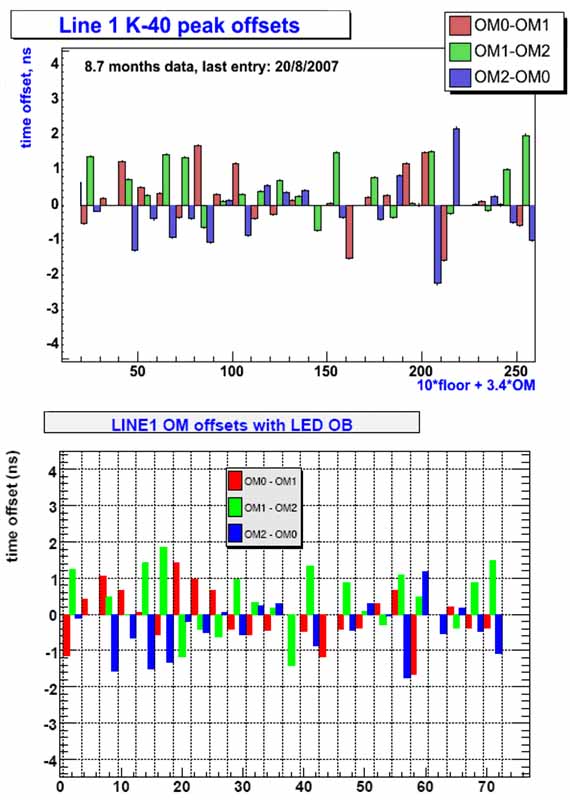}
   \caption{Time difference between three PMTs located in the same floor in line 1 computed with the K40 (up) and with the Optical Beacons (down), where X axis represents number of OM. In the ideal case (perfect calibration) the bars should be essentially zero.}
   \label{simp_fig}
 	\end{figure}

\item The Laser Beacon system is a device capable of emitting intense ($\sim$ 1 $\mu$J) and short ($<$1 ns) light pulses (see Fig. 4). In ANTARES there are two LOB placed at the bottom of two central lines from where are able to illuminate most of the PMTs of the detector. This system can be used for interline time calibration of the OMs in the lower floors of the detector. In addition it can be used to perform cross-check with positioning systems of the detector by computing the time difference, for a given time period, between the LOB emission and the recording in the OMs of some particular floors on a line. Proceeding in this way we could see that the distribution of the values improves significantly when the real shape of the lines provided by the positioning system is considered (RMS $\sim$0.6 ns), rather than thinking in rigid straight lines (RMS $\sim$2.3 ns).

\end{enumerate} 	

\subsection{Potassium-40}
The potassium-40 is a $\beta$-radioactive isotope naturally present in the sea water. The decay of this substance produces an 1.3 MeV energy electron that will exceeds the Cherenkov threshold inducing a light cone capable of illuminate two OMs in coincidences if the emission occurs in the vicinity of a detector floor.  This event will result on a visible bump over the distribution of the relative time delays between hits in two PMTs of the same floor. Ideally, this coincidence peak must be centered at zero position.  Therefore, the K40 offers a completely independent method to compute the time offsets [5]. Experimental measurements show a small spread of the offsets in good agreement with those values obtained with the Optical Beacons systems (see Fig. 5), and confirm the  accuracy of the timing calibration in ANTARES.\\\section{Conclusions} 
In this paper I have reviewed the different methods used in the ANTARES Neutrino Telescope to perform the timing calibration. The results from laboratory an in-situ measurements confirm their validity and the high level of accuracy reached. With regard to the absolute time resolution the demanded resolution of $\sim$1 ms could be reached by the precision of the clock system. The time resolution was found to be $<$0.5 ns by the Optical Beacons systems. This result was validated by the independently measuments from the K40 coincidences analysis.\\

\end{document}